# Role of a polymeric component in the phase separation of ternary fluid mixtures: A dissipative particle dynamics study

Amrita Singh[a], Anirban Chakraborti[a], and Awaneesh Singh[b]

[a] School of Computational and Integrative Sciences, Jawaharlal Nehru University, New Delhi-110067, India.

[b] Department of Physics, Institute of Chemical Technology, Mumbai-400019, India. E-mail: awaneesh11@gmail.com

**Abstract**

We present the results from dissipative particle dynamics (DPD) simulations of phase separation dynamics in ternary (ABC) fluids mixture in $d = 3$ where components A and B represent the simple fluids and component C represents a polymeric fluid. Here, we study the role of polymeric fluid (C) on domain morphology by varying composition ratio, polymer chain length, and polymer stiffness. We observe that the system under consideration lies in the same dynamical universality class as a simple ternary fluids mixture. However, the scaling functions depend upon the parameters mentioned above as they change the time scale of the evolution morphologies. In all cases, the characteristic domain size follows: $\ell(t) \sim t^\phi$ with dynamic growth exponent $\phi$, showing a crossover from the viscous hydrodynamic regime ($\phi = 1$) to the inertial hydrodynamic regime ($\phi = 2/3$) in the system at late times.

## 1. Introduction

Much recent interest has been focused on the kinetics of phase separation of homogeneous multicomponent mixtures, which are rendered thermodynamically unstable by a rapid temperature quench below the critical temperature of the system.[1–3] After the quench, the system starts evolving towards equilibrium via the formation and growth of domains of different ordered phases. This far-from-equilibrium coarsening process has been studied extensively, binary (AB) mixtures in particular, by using various simulation methods, experimentally, as well as analytically.[4–6] It is now well established that the evolving system segregates into A-rich and B-rich domains with a characteristic length scale $\ell(t)$, where $t$ is the time after the quench. The existence of a characteristic length scale during the late times leads to the dynamical scaling exhibited by the two-point equal-time correlation function $C(r, t)$ and its Fourier transform, the structure factor $S(k, t)$,

$$C(r, t) = g(r/\ell(t)),$$
$$S(k, t) = \ell(t)^d f(k\,\ell(t)). \qquad (1)$$

Here, $g(x)$ and $f(p)$ are scaling functions; $r$ is the spatial separation between two points; $k$ is the magnitude of the wave vector, and $d$ is the dimensionality of the system.[7]

The length scale $\ell(t)$ is related to average domain size and displays a simple power law dependence with time: $\ell(t) \sim t^\phi$, where $\phi$ is the growth exponent, and it depends on the transport mechanism which drives the segregation, *e.g.*, $\phi = 1/3$ for diffusive dynamics in $d \geq 2$ and it is the only growth exponent expected for phase separating binary alloys.[1,2] However, in the case of simple fluids mixture and polymeric fluid mixtures, hydrodynamic effects appear at later times that lead to the various growth regimes characterized by different growth exponents. For example, in $d = 3$ phase separating fluids, convective transport mechanism results in following growth regimes[8–10] with

$$\phi = 1/3, \quad \ell(t) \ll D\eta^{1/2}, \quad \text{(diffusive regime)}$$
$$\phi = 1, \quad \ell(t) \ll \ell_{in}, \quad \text{(viscous hydrodynamic regime)}$$
$$\phi = 2/3, \quad \ell(t) \gg \ell_{in}. \quad \text{(inertial hydrodynamic regime)} \qquad (2)$$

Here, the inertial length $\ell_{in} \approx \eta^2/(\rho\lambda)$ where, $\eta$, $\rho$, and $\lambda$ are the viscosity, density and surface tension, respectively; D is the diffusion coefficient.

Few experimental studies have reported the evidence for a crossover from diffusive to viscous hydrodynamic regime.[11–13] However, the observation of an inertial regime has not been reported by any experiment till date.[11-13] The viscous hydrodynamic regime has been observed in some numerical simulations which include the studies of *Model H* or its variants.[14–16] Both viscous and inertial regimes have been finally observed only by *Lattice-Boltzmann* simulations,[17,18] which are analogous to coarse-grained phenomenological models. Later, in a microscopic-level large-scale molecular dynamic (MD) simulation studies, linear growth law in the viscous hydrodynamic regime is confirmed.[19,20] However, due to high computational cost, MD simulations have not yet accessed the inertial hydrodynamic regime for $d = 2$ and $d = 3$ systems.

Though there is a good understanding of phase separation dynamics in binary (AB) mixture, yet our understanding of domain growth kinetics is somewhat limited for segregating ternary fluid mixtures (ABC); this is because in ABC system, the phase ordering competition results in a more complex problem. To provide a proper context to our work, let us have a brief review of some of the studies on phase separating ternary (ABC) fluid mixtures. In an MD study of segregation kinetics in ABC fluids in $d = 2$ Laradji *et al.*[21] found that the hydrodynamic flow did not control the coarsening as their results suggested the growth law: $\ell(t) \sim t^{1/3}$; but, their simulation cannot genuinely be referred to as late-stage study as they could only access the diffusive growth regime ($\phi = 1/3$).

Further, in the hydrodynamic lattice-gas simulations of AB and symmetric ABC fluids, in $d = 2$, Lakshami *et al.*[22] observed the early-times diffusive growth regime ($\phi = 1/3$) and a viscous regime ($\phi = 1/2$) at late times. They also recovered the inertial growth exponent, $\phi = 2/3$, after reducing the volume fraction of one of the components. In a recent MD simulation study, Singh *et al.*[23] have confirmed the crossover from $\phi = 1/3 \to \phi = 1/2 \to \phi = 2/3$ in $d = 2$. They have also accessed the viscous hydrodynamic regime $\phi = 1/3 \to 1$ in $d = 3$. Till date, no evidence of a crossover from the viscous hydrodynamic regime ($\phi = 1$) to the inertial hydrodynamic regime ($\phi = 2/3$) for $d = 3$ phase separating ternary fluids has been provided by any MD simulation study.

There have been several experimental studies of phase separating polymeric fluid mixtures[24–27] demonstrating that the systems of polymer mixtures of the components with dynamical symmetry, lie in the same dynamical universality class as that of simple binary or ternary fluids, i.e., they exhibit the same growth regimes as in Eq. (2). On the contrary, in some of the experimental works, Tanaka[28,29] has argued that a phase separating polymeric fluids mixture, when there is a dynamical asymmetry between the components shows anomalously slow growth and such systems do not belong to the same universality class as the *Model H*.

Later on, some MD studies of polymeric fluid segregation process in $d = 2$ and $d = 3$ have accessed the early time diffusive growth regime ($\phi = 1/3$) but they also could not go beyond the early-time diffusive regime to access the hydrodynamic growth regime.[30,31] In one more large-scale MD simulation studies of segregating polymer-solvent mixture, where there was a dynamical asymmetry between the components as the solvent molecules were obviously more mobile than the polymers, Virnau *et al.*[32] could not recover the viscous growth or inertial growth regimes, on the time scale of their simulations. Finally, in a more recent MD study, Singh *et al.*[33] had demonstrated that the segregating polymeric mixtures with dynamical symmetry lie in the same universality class as that of simple binary fluids. They also observed the crossover from diffusive $\phi = 1/3$ to viscous $\phi = 1/2$ to inertial growth regime $\phi = 2/3$ in $d = 2$ and a crossover from $\phi = 1/3 \to 1$ in $d = 3$. To the best of our knowledge, there is no evidence of a crossover from viscous ($\phi = 1$) to inertial hydrodynamic regime ($\phi = 2/3$) in the case of segregating polymer mixtures in $d = 3$, as far as MD results are concerned.

In this work, we study the kinetics of phase separation in a ternary fluids mixture (which has two simple fluid components and one polymeric fluid) using Dissipative Particle Dynamic (DPD) simulations[34,35] in $d = 3$. Our DPD results lead to the first unambiguous confirmation of a crossover from the viscous hydrodynamic regime to the inertial growth regime for binary (AB) and ternary (ABC) fluid mixtures (where component C is considered as a simple fluid in one case and a polymeric fluid in another case). We further investigate, the effect of polymeric fluid (i.e., C-component of ABC fluid) on domain morphology. In particular, we study how the domain morphology is affected by varying some parameters such as polymeric fluid concentration, polymer chain length, and polymer chain stiffness.

This paper is organized as follows. In Sec. 2, we describe our model and methodology with simulation details. We present our DPD simulation results and discussions in Sec. 3. Finally, Sec. 4 concludes this paper with a summary.

## 2. Computational method and simulation details

Dissipative particle dynamics (DPD) is a powerful simulation technique used to model the dynamic behavior of the evolving complex fluids.[34,35] In general, DPD can be viewed as a coarse-grained molecular dynamics (MD) approach where individual beads represent molecules or clusters of molecules interacting through a soft-core potential. This makes DPD more useful technique to simulate more substantial length and time scales than traditional MD simulations where individual atoms interact through a hard-core Lennard-Jones potential. Notably, it is feasible to perform simulations of polymeric fluids in volumes up to 100 *nm* in linear dimension and for time scales of tens of microseconds.[34,35]

The time evolution of the system is governed by integrating Newton's equation of motion: $m\, d\boldsymbol{v}_i/dt = \boldsymbol{f}_i$. The effective force $\boldsymbol{f}_i(t)$ acting on each bead $i$, consists of three pairwise additive forces which are conservative force ($\boldsymbol{F}_{ij}^C$), dissipative force ($\boldsymbol{F}_{ij}^D$), and random force ($\boldsymbol{F}_{ij}^R$) such that $\boldsymbol{f}_i(t) = \sum \boldsymbol{F}_{ij}^C + \boldsymbol{F}_{ij}^D + \boldsymbol{F}_{ij}^R$. Here, the sum runs over all beads $j$ within a cut-off radius $r_c$ from bead $i$. The conservative force acting between beads $i$ and $j$ separated by a distance $r_{ij} = |\boldsymbol{r}_i - \boldsymbol{r}_j|/r_c$ is a soft repulsive interaction which is linear up to the cutoff distance $r_c$ and is given by $\boldsymbol{F}_{ij}^C = a_{ij}(1 - r_{ij})\hat{\boldsymbol{r}}_{ij}$. Here, $a_{ij}$ measures the maximum repulsion between beads and $\hat{\boldsymbol{r}}_{ij} = \boldsymbol{r}_{ij}/|\boldsymbol{r}_{ij}|$ determines the direction of the force. The dissipative force is $\boldsymbol{F}_{ij}^D = -\gamma\omega_D(r_{ij})(\hat{\boldsymbol{r}}_{ij} \cdot \boldsymbol{v}_{ij})\hat{\boldsymbol{r}}_{ij}$, where $\gamma$ is the strength of the dissipation (friction coefficient) between beads, $\omega_D$ is a weight function that goes to zero at $r_c$, and $\boldsymbol{v}_{ij} = \boldsymbol{v}_i - \boldsymbol{v}_j$.

The random force is $\boldsymbol{F}_{ij}^R = \sigma\omega_R(r_{ij})\xi_{ij}\hat{\boldsymbol{r}}_{ij}$, where $\sigma$ is the noise amplitude, $\xi_{ij}$ represents a Gaussian random variable of unit variance and zero-mean: $\langle\xi_{ij}(t)\rangle = 0$ and $\langle\xi_{ij}(t)\xi_{kl}(t')\rangle = (\delta_{ik}\delta_{jl} + \delta_{il}\delta_{jk})\delta(t - t')$. The symmetry property $\xi_{ij} = \xi_{ji}$ ensures the local momentum conservation. The simulation samples a canonical ensemble and obeys the fluctuation-dissipation theorem where the noise amplitude, and the friction coefficient are related as $\sigma^2 = 2\gamma k_B T$; $k_B$ is the Boltzmann constant, and $T$ is the temperature of the system. The weight functions are chosen of the following form[34,35]: $\omega_D(r_{ij}) = \omega_R(r_{ij})^2 = (1 - r_{ij})^2$ for $r_{ij} < r_c$ to ensure the correct canonical enesemble. The value of $\gamma = 4.5$ is chosen to ensure relatively rapid equilibration of the temperature of the system and the numerical stability of the simulations for the specified time step.[24,36–38] Each of these three pair wise forces conserves the momentum locally, and thus, DPD simulations reproduce correct hydrodynamic behavior even in systems containing only a few hundred particles.[34,35]

The time evolution of the system is captured by integrating the equation of motion via a modified velocity-Verlet algorithm.[39] In the simulation, $r_c$ is taken as the characteristic length scale and $k_B T$ as the characteristic energy scale. We use the dimensionless value of $r_c = 1$. In our simulations, we quench the system at a reduced temperature $T^* = T/T_0 = 1$, where $T_0$ is taken as a reference temperature. As we show subsequently, this quench temperature is well below the corresponding critical temperature for phase separation. The interaction parameter $a_{ij}$ is given in terms of $k_B T_0/r_c$; the characteristic time scale is defined as $\tau = (mr_c^2/k_B T_0)^{1/2}$. The system is integrated at the time step $\Delta t = 0.02\tau$. The total bead number density is $\rho = 3$, which is a reasonable choice for DPD simulation of liquids.[34,35] The dimensional values of length $r_c$ and time $\tau$ are observed to be 0.97 *nm* and 8.3 *ps*, respectively.[40,41]

The polymer chains are simulated using a bead-spring model, with the harmonic bond potential given by $E_b = 1/2\, k_b (r - r_0)^2$. Here, $k_b = 128$ is the elastic constant and $r_0 = 0.5$ is the equilibrium bond distance.[36,37,42,43] The stiffness of polymer chain is dominated by the angle potential: $E_a = 1/2\, k_a (\cos\theta - \cos\theta_0)^2$, where $\theta$ is the angle between the sequential bonds along a chain itself, $k_a$ is the angle coefficient, and $\theta_0 = 180$ is the equilibrium value of the angle.[44] In our coarse-grained simulation approach, all the molecules are modeled as DPD beads. The repulsive interaction parameter between any two beads of the same component (compatible) is set to $a_{ij} = 25$ (in units of $k_B T/r_c$), and that between the two beads of different components (incompatible), is set to $a_{ij} = 50$.[34,35] The choice of our interaction parameters energetically favors phase separation in the system.

The size of our simulation box is $L_x \times L_y \times L_z = 64 \times 64 \times 64$ units with periodic boundary conditions in all three directions. The total number density of beads in the simulation box is set to $\rho = 3$, hence the total number of beads in the system is $N = 3 \times 64^3 = 786432$. This high density ensures that the system is far away from the gas-liquid transition.

At the onset of DPD simulation, we perform a comparative study of segregation kinetics of simple binary (AB) fluids with the critical composition $N_A:N_B = 1:1$ and simple ternary (ABC) fluids with three different ratios of $N_A$, $N_B$, and $N_C$ which are $1:1:1$ (critical composition), $2:2:1$, and $4:4:1$ (off-critical composition), respectively; $N_A$, $N_B$, and $N_C$ are the number of A-type, B-type, and C-type of beads in the system such that $N = N_A + N_B + N_C$. Further, we consider a ternary mixture which consists of two simple fluid components (A and B) and a polymeric fluid C. Our purpose of studying phase separation in the polymeric fluid mixed ternary mixture is multifold: (1) we study the domain morphologies for three different polymer chain lengths ($L_p = 16, 32,$ and $64$) for each ratio of $N_A$, $N_B$, and $N_C$ as mentioned above. The number of polymer chains ($N_{pc}$) of length ($L_p$) in the system is defined as: $N_{pc} = N_C/L_p$. (2) We investigate the effect of polymer chain length, composition and polymer stiffness on the domain morphologys and master scaling function. Particularly, we perform DPD simulations for three values of $k_a: 5, 20$ and $60$, each with above mentioned $L_p$'s and composition ratios.

### 3. Results and discussion

In this section, we present detailed numerical results for phase separation kinetics in ternary fluid mixtures with various composition ratios of A, B, and C-type of beads and at different polymer chain lengths and stiffness.

### 3.1 Characterization of evolution morphologies

As already stated, domain coarsening is a scaling phenomenon for the system under consideration where morphologies at two different times are self-similar. Among various morphology characterization functions, two most important ones are the two-point equal-time order parameter correlation function $C(r,t)$ and its Fourier transform, the structure factor $S(k,t)$. For ternary fluids, we define two types of correlation functions. The first one is given by:

$$C(r,t) = \langle \psi(r_1,t)\psi(r_2,t) \rangle - \langle \psi(r_1,t) \rangle \langle \psi(r_2,t) \rangle, \tag{3}$$

where $r = r_1 - r_2$. $\psi(r_1,t)$ is the order parameter (which is the local concentration difference of various components in the mixture) at a discrete site $r_1$ at time $t$. The angular brackets represent an ensemble average, which is obtained over five independent runs. This correlation function refers to the morphologies of A and B-type of beads. The second correlation function is defined for the order-parameter $\psi(r_1,t)^2$, which signifies the domain morphology of C-beads[45]:

$$D(r,t) = \langle \psi(r_1,t)^2 \psi(r_2,t)^2 \rangle - \langle \psi(r_1,t)^2 \rangle \langle \psi(r_2,t)^2 \rangle. \tag{4}$$

The structure factor is computed by taking the Fourier transform of $C(r,t)$ as

$$S_C(k,t) = \int dr\, e^{ik.r} C(r,t), \tag{5}$$

where $k$ is the scattering wave-vector. Similarly, we represent the Fourier transform of $D(r,t)$ as $S_D(k,t)$.

We obtain the order parameter $\psi(r,t)$ by considering non-overlapping boxes of unit size. Then the continuum fluid configurations are mapped onto a simple cubic lattice of size $64 \times 64 \times 64$. We now count the number of A, B, and C-beads in each box. If a box is effectively occupied by the A-type of beads, we assign $\psi(r,t) = +1$. Similarly, a box, occupied by the B-type of beads is assigned $\psi(r,t) = -1$, and that occupied by C-type of beads is assigned $\psi(r,t) = 0$. Boxes with the equal number of beads are assigned $\psi = +1$ or $\psi = -1$ or $\psi = 0$ with equal probabilities. In a similar way, we assign $\psi(r,t) = +1$ and $-1$ for A and B-types of beads, respectively, for the binary fluid mixture.[19,20,23]

It is notable that the domain structure is visible from the higher-order Bragg-reflections in the structure factor, namely at $(2n+1)k_m$ for $n = 1, 2, \cdots$. The scattering around $k_m$ dominates the structure factor spectrum and leads to narrow peak. Since we are working in a finite lattice, only a discrete set of $k's$ is physically meaningful: $k = (2\pi n/L)$ with $n = (n_x, n_y, n_z)$ where $0 \leq n_i \leq L$ for $i = x, y, z$. The separated phase ordering is then detected by the *number of peaks* in the structure factor produced by the periodicity of the morphology for all these $k's$. Since our system is isotropic, statistics of both the correlation functions and the structure factors can be improved by taking the spherical averaging. We represent corresponding spherically averaged quantities by $C(r,t)$, $S_C(k,t)$, $D(r,t)$, and $S_D(k,t)$, respectively.

The foremost consequence of the existence of a characteristic length scale $\ell(t)$ is that the system demonstrates scaling behavior as shown in Eq. (1). The average domain size $\ell(t)$ can be computed by exploiting the scaling properties of various morphology characterization functions. In this work, we define $\ell(t)$ as the distance at which the correlation function decays to some fraction of its maximum value ($C(0,t) = 1$, and $D(0,t) = 1$). We observed that the first zero-crossing of correlation function (note that the correlation function displays damped oscillation before decaying to zero in conserved kinetics) gives an appropriate measure of the average domain size $\ell_C(t)$ and $\ell_D(t)$, respectively. Several other suitable definitions for computing $\ell(t)$ are: half-crossing of the correlation function, the inverse of the first moment of $S(k,t)$, and finally, the first moment of the normalized domain-size distribution function. However, all these definitions differ only by constant multiplicative factors.[46,47]

### 3.2 Simple binary and ternary fluid mixtures

To begin with, we investigate the previously studied cases of segregating simple binary and ternary fluids where the asymptotic inertial growth regime was not observed in $d = 3$.[23] The reason could be the smaller system size used ($N = 110592$, and number density $\rho = 1$)[23] or the technique that could preserve the hydrodynamics significantly better even in the smaller system. Therefore, we perform DPD simulations for a system of a large number of beads, $N = 786432$ in a box of size $64 \times 64 \times 64$ and density $\rho = 3$. Here, DPD simulations are best known for preserving the hydrodynamics even with a few hundreds of particles in the system as far as particle-based simulations are concerned.[34,35]

The initial homogeneous configuration in this case, and for all the other cases discussed next, are prepared at a high-temperature $T = 5$ (configuration of DPD particles is generated randomly) and then further equilibrated for $t = 200$.[23,33] Then we quench the system from a high temperature homogeneous phase to a lower temperature $T = 1$ at $t = 0$. After the quench, we monitor the evolution of the system at various times.

We present the evolution of segregating ternary fluid mixtures in Fig. 1 using isosurface plots corresponding to different composition ratios A:B:C = 1:1:1 (critical/symmetric mixture) in 1a, 2:2:1 in 1b, and 4:4:1 in 1c, respectively. These snapshots are obtained at time $t = 200$ after the quench. Fig. 1d displays the isosurface plot of the critical binary mixture (A:B = 1:1) at the same time. For the ternary fluids, we clearly see the emergence of A-rich, B-rich, and C-rich domains depicted by open space, blue patches, and green patches, respectively, with all the three interfaces namely AB, BC, and CA. The morphology in the symmetric ternary mixture (Fig. 1a) where all the beads are in equal proportions is termed as "blob" morphology.[45] Further, for the asymmetric ternary mixtures (A:B:C = 2:2:1 and 4:4:1), where component C is decreasing with corresponding increase in A and B components, we observe a sign of formation of bicontinuous domain structures between A- and B-rich phases (Fig. 1b,c). With the reduction of component C to zero (i.e., A:B:C = 1:1:0), which corresponds to a symmetric binary (AB) mixture, a usual bicontinuous domain structure is seen (Fig. 1d). However, only for the isosurface plot, we map $\psi(r,t)$ value in the range 1 to 3 (A $\in$ (0, 1), B $\in$ (1, 2), and C $\in$ (2, 3)) to differentiate the colors of interfaces (see the color bar).

We calculate the radial distribution function (RDF), $g_{AB}(r)$ to characterize the spatial distribution of A and B-beads. Fig. 2a represents the comparison of RDF at time $t = 200$ for different ratios of A:B:C corresponding to the snapshots shown in Fig. 1. To calculate $g_{AB}(r)$, we compute the probability of an A-type bead being at a distance, $r_{ij} = \left(\left|dx_{ij}\right|^2 + \left|dy_{ij}\right|^2 + \left|dz_{ij}\right|^2\right)^{1/2}$, from a B-type bead. Since our system is periodic in all the three directions, we assume that the maximum value of $dx_{ij}$, $dy_{ij}$, and $dz_{ij}$ is less than or equal to half of the periodic box length.[48] As the proportion of C-beads in the system decreases (corresponding proportion of A and B-beads also increases), the height of the peak for $g_{AB}(r)$ clearly increases and the peak position shifts to larger value of $r$. The lower peak of $g_{AB}(r)$ at small $r$ for a symmetric ternary mixture (black curve) indicates fewer A-beads in the proximity of B-beads (due to the presence of C-beads). Hence, the size of A and B domain are smaller than the symmetric binary (AB) mixture (blue curve) where the high peak of $g_{AB}(r)$ at large $r$ indicated that A-beads are closely clustered around B-beads with large cluster sizes.

We show the comparison of the scaling functions (Figs. 2b and 2c) to discuss whether the domain morphology depends on the degree of off-criticality for the evolution depicted in Figs. 1a-d, at time $t = 200$.

We plot data for $C(r,t)$ vs. $r/\ell_C$ in Fig. 2b, where $\ell_C$ denotes the first zero-crossing of $C(r,t)$. A horizontal black solid line is drawn to enhance the zero crossing of $C(r,t)$. The scaled correlation function for ternary mixtures for three different ratios is denoted by the specified symbol types. The solid curve refers to the scaled correlation function for binary mixture (A:B:C = 1:1:0; corresponding evolution is depicted in Fig. 1d). A reasonably good data collapse indicates that the evolving systems are the part of the same dynamical universality class. This is also noticeable from the fact that in the calculation of the correlation function, order parameter, $\psi(r,t) = 0$, does not contribute. Therefore, the scaling behavior of $C(r,t)$ for ternary fluid mixture and that for a binary fluid should be comparable. A log-log plot of the scaled structure factor, $S_C(k,t)\ell_C^{-3}$, as a function of $k\ell_C$ (see the inset) for the same data set is further demonstrating the dynamical scaling. For a reference, the scaled structure factor data for a binary mixture, denoted by a solid black line is also included.

The scaling plot of $D(r,t)$ vs. $r/\ell_D$ (at $t = 200$, for three different composition ratios as indicated by three symbols) demonstrates that the scaled correlation function, $D(r,t)$, clearly depends upon the composition ratios (A:B:C) (see Fig. 2c). Recall the definition of $D(r,t)$, which is equivalent to the correlation function for a binary fluid mixture ($\psi^2 = 0, 1$) with off-critical composition and it is a well-known fact that the correlation function varies continuously with the degree of off-criticality.[2,49] The corresponding scaled structure factor ($S_D(k,t)\ell_D^{-3}$ versus $k\ell_D$) plot further confirms that data for different ratios do not collapse onto a single master curve; the deviation is more visible at smaller $k$ values (see the inset of Fig. 2c). However, the power law decays ($\sim k^{-4}$) of the tails for both $S_C(k,t)$ and $S_D(k,t)$ are consistent with the expected Porod's law, $S(k,t) \sim k^{-(d+1)}$, for $k \to \infty$, ($d$ is the system dimensionality), which results from the scattering off sharp interfaces.[50,51]

To investigate the time dependence of domain size, we plot $\ell_C(t)$ *vs.* $t$ on a logarithmic scale (see Fig. 2d). Here, and in all successive discussions we quantify the growth as $\ell_C(t) = \ell_0 + \alpha t^\phi$ where $\ell_0$ and $\alpha$ are temperature dependent quantities, so remain constant. Therefore, we plot the average domain size, $\ell_C \equiv \ell_C - \ell_0$ and time, $t \equiv t - t_0$. The length $\ell_0$ is the average domain size of the transient regime, when the system becomes unstable due to fluctuations at early time $t_0$ since the quench.[52,53] Here, $t_0 = 4$ is the onset time for different composition ratios and their corresponding $\ell_0$ values are 1.454 (for A:B:C = 1:1:1), 1.538 (for A:B:C = 2:2:1), 1.576 (for A:B:C = 4:4:1), and 1.621 (for A:B:C = 1:1:0), respectively.[53] We notice that the diffusive regime ($\phi = 1/3$) is very short lived on the time scale of our simulation, as DPD simulation is best known for preserving the hydrodynamics behavior of the system.[34,35] A clear crossover is observed from the viscous hydrodynamic regime ($\phi = 1$) to the inertial hydrodynamic regime ($\phi = 2/3$). The expected growth exponents in various growth regimes are represented by the dashed lines. In the case of symmetric ternary (A:B:C = 1:1:1) mixture, the evolution time-scale appears to be a bit slow. Reason being the lack of connectivity between the domains which consequently affects the hydrodynamic transport. While in asymmetric ternary mixtures, the number of A and B-beads are increased at the cost of C-beads, which results in the increased connectivity between A-rich and B-rich domains.

### 3.3 Ternary mixture with one polymeric component

Now we focus our attention on phase separation dynamics in a ternary (ABC) fluids mixture where a covalent-bonded polymeric fluid replaces C-component. At time $t = 0$, we quench the homogeneous system and follow the evolution at different times. Recall that our focus is to investigate the effect of polymeric fluid component on domain morphologies and the scaling functions of the mixtures.

### 3.3.1 Varying composition ratio in the mixture

First, we consider the case where polymer chain length ($L_P = 32$) and polymer stiffness due to angle-potential ($k_a = 20$) are fixed. Then, we vary the composition ratios ($N_A: N_B: N_C$) and compare the corresponding domains morphologies at different times (see Fig. 3). In Figs. 3a-c, we present evolution snapshots (isosurfaces) obtained from our DPD simulations for a symmetric (A:B:C = 1:1:1) and two asymmetric (A:B:C = 2:2:1, 4:4:1) mixtures at $t = 100, 500$. The snapshots in Fig. 3a clearly demonstrate the formation of A-, B- and C-rich domains as indicated by open space, blue, and green patches, respectively, with all three interfaces AB, BC, and CA present. To verify the coarsening of these domains, we plot the spatial number density distributions for A-beads ($\rho_A(z)$) along the transverse ($z$) direction in Fig. 3d at $t = 100$ (black circle), 300 (red square), and 500 (green triangle), respectively. The increase in height and width of the number density curve with time confirms the evolution of A-rich domains. The evolution patterns corresponding to other two asymmetric (off-critical) compositions, 2:2:1 (Fig. 3b) and 4:4:1 (Fig 3c), exhibit the similar behaviors and patterns as in Fig. 3a. However, the comparison of density profile curves exhibits that with the increase of off-criticality, the system becomes more segregated, *i.e.*, the average size of A-rich domains in Fig. 3c is larger than in Fig. 3a. This feature is similar to that of segregating simple ternary mixtures.

To further characterize the spatial distribution of domains within the system, we calculate the radial distribution function, $g_{AB}(r)$, of A- and B-beads at $t = 500$ for different composition ratios plotted in Fig. 4a. This plot shows that the peak of $g_{AB}(r)$ is enhanced and shifted to the higher value of $r$ with the increase in off-criticality, which complements our results in Fig. 3.

To examine the statistical properties of the evolution depicted in Fig. 3, we calculate the order parameter, $\psi(\mathbf{r}, t)$, using the same approach as discussed in Sec. 3.2. The results are averaged over 5 independent runs. In Fig. 4b, we show a comparison of the scaled correlation function ($C(r, t)$ vs. $r/\ell_C$) and the corresponding structure factor ($S_C(k, t) \ell_C^{-3}$ vs. $k\ell_C$ on a log-log scale) for three different compositions as indicated by different symbols at $t = 500$, when the system is already in the scaling regime. A neat data collapse suggests that they belong to the same dynamical universality class. One can expect this behavior as $\psi(\mathbf{r}, t) = 0$ makes no contribution in the calculation of the scaling function and signifies, particularly, the morphologies of A and B-types of beads. Therefore, the scaling behavior of $C(r, t)$ and $S_C(k, t)$ of a ternary mixture with a polymeric component retains the same universality class as the simple ternary and binary fluid mixtures. The structure factor tail obeys the Porod's law, $S_C(k, t) \sim k^{-4}$ at large $k$ values for all compositions. In Fig. 4c, we demonstrate the scaling behavior of the correlation function, $D(r, t)$ vs. $r/\ell_D$, and the structure factor, $S_D(k, t) \ell_D^{-3}$ vs. $k\ell_D$, that emphasizes the morphology of polymeric fluid (C-component). The non-overlapping curves in both the plots indicate that the scaling functions clearly depend upon the composition ratios of A, B, and C-beads. However, we observe that the deviation of the scaling functions, $D(r, t)$, from the master curve at different ratios, for a ternary mixture with a polymeric component (Fig. 4c) is more significant than the simple ternary and binary fluid mixtures (Fig. 3c).

The time dependence of average domain size (for the evolution shown in Fig. 3) is presented in Fig. 4d, where we show a log-log plot of $\ell_C(t)$ *vs.* $t$ for the different composition ratios indicated by different symbol types. Here, again we exclude the transient regime data at onset time $t_0 = 4$. We do not see the short-lived diffusive regime ($\phi = 1/3$) at the time-scale of our simulation and the system quickly enters the viscous hydrodynamic regime ($\phi = 1$). At late times, the domain growth gradually evolves towards the inertial growth regime ($\phi = 2/3$) which is qualitatively similar to the simple fluid mixtures depicted by the blue line. Moreover, we observe that with increased asymmetry in the system, evolution time-scale appears to be faster firstly, due to the increased connectivity between A- and B- beads as the number of C-beads decreases (with the corresponding increase in A- and B-beads) that makes the hydrodynamic transport more effective along the interface. Secondly, the diffusion coefficient of simple fluid is higher than polymeric fluid and therefore the higher polymer concentration (C-beads) in the mixture slows down the evolution process.[54]

### 3.3.2 Varying polymer chain length in the mixture

Let us next discuss the effect of polymer chain length ($L_p$) variation on the domain coarsening. For this, we consider the symmetric ternary (A:B:C = 1:1:1) mixture having polymer chains (C-component) of same stiffness ($k_a = 20$), but three different lengths, $L_p = 16, 32,$ and $64$.

First, we compare RDF of A-beads around B-beads for different chain lengths at $t = 500$ for the evolution shown in Fig. S1. The peak position of RDF curves (see Fig. 5a) increases slightly with a notable shift to smaller $r$ values, confirming the smaller domain size of A- or B-type of beads as the chain length increases from $L_p = 16 \rightarrow 64$. This is due to the fact that the diffusion coefficient of polymer depends on the degree of polymerization ($L_p$) as $D \sim L_p^{-\nu}$, where $\nu = 1/2$ is the size-exponent of a polymer in the melt, and $\nu = 2/3$ in dilute solution.[54,55] This indicates that, at a fixed time, a system with shorter chain length is the one with larger average domain size. This is also apparent from the evolution snapshots in Fig. S1 (see Supplementary Information (SI)).

To understand the statistical properties of the evolution morphology shown in Fig. S1, we plot the scaled correlation functions, $C(r,t)$ vs. $r/\ell_C$ in Fig. 5b, and the corresponding structure factors, $S_C(k,t)\,\ell_C^{-3}$ vs. $k\ell_C$ in the inset on a logarithmic scale. The plot is for three different values of $L_p$ at $t = 500$ as indicated by different symbols. The excellent data collapse in both the plots demonstrates that the systems with different chain lengths belong to the same dynamical universality class. This is again expected as in the calculation of $C(r,t)$, the contribution of order parameter corresponding to C-beads is zero (discussed earlier). Fig. 5c, shows the scaling behavior of $D(r,t)$ vs. $r/\ell_D$ with the inset log-log plot of the corresponding structure factor $S_D(k,t)\,\ell_D^{-3}$ vs. $k\ell_D$. A noticeable deviation is observed between the curves from the master function in both the plots. For $S_D(k,t)$, this deviation is more distinct at smaller $k$ values (i.e., for larger domain sizes). This indicates that the scaling functions of segregating ternary mixtures with polymeric fluid as one of the component depend on the polymer chain length. However, for a phase separating binary (AB) polymer mixture (where A and B both are polymers), the scaling functions are independent of the degree of polymerization.[33]

Further, we compare the time dependence of average domain size for the three different values of $L_p$ indicated by the respective symbols in Fig. 5d. A length scale for the simple ternary critical fluid mixture (A:B:C = 1:1:1), represented by a blue line, is also displayed for the comparison. Here, we plot, $\ell_C(t)$ vs. $t$, on a log-log scale. Notice that, as $L_p$ increases, evolution time-scale becomes slower because of the slower migration of longer chains that causes poor connectivity between domains of A-type and B-type of beads. Therefore, the evolution time-scale for simple ternary mixture is higher than that of the ternary mixture with one polymeric component. Our data sets clearly show a crossover from the viscous hydrodynamic regime to the inertial hydrodynamic regime ($t^1 \rightarrow t^{2/3}$) which are represented by the dashed lines of slope 1 and 2/3, respectively.

### 3.3.3 Varying polymer stiffness ($k_a$) in the mixture

Finally, we compare the evolution morphologies of ternary fluid mixtures with symmetric composition (A:B:C=1:1:1), for three different values of polymer chain stiffness ($k_a = 5, 20$ and $60$) at fixed chain length $L_p = 32$. To show the comparison, we plot the evolution snapshots in Fig. S2 (see in SI) and the statistical properties of the system in Fig. 6.

Fig. 6a displays the comparison of RDF's of A-beads about B-beads at $t = 500$, for three values of $k_a$ as indicated by the different symbol types. The peak of $g_{AB}(r)$ shifts towards lower $r$ values as we increase the polymer chain stiffness, *i.e.*, increasing $k_a$ leads to the less segregated system, or a system with smaller average domain size. This is also evident from the corresponding evolution pictures in Fig. S2.

We next probe the effect of polymer chain stiffness on the coarsening process by comparing the scaled correlation functions $C(r,t)$ *vs.* $r/\ell_C$ and $D(r,t)$ *vs.* $r/\ell_D$, for three different values of $k_a$ at $t = 500$ in Figs. 6b and 6c, respectively. In the inset, we plot the corresponding structure factor data sets. As expected, the neat overlapping of data sets (in Fig. 6b) demonstrates that the system, where the order parameter of C-beads ($\psi(r,t) = 0$) does not contribute in the calculation of the correlation function, lies in the same dynamical universality class as that of the simple ternary fluid mixtures. On the other hand, Fig. 6c shows a slight deviation of the curves from the master function in both $D(r,t)$ and respective $S_D(k,t)$; this confirms the dependency of scaling functions on the polymer chain stiffness ($k_a$). However, the deviations of the scaling functions from the master curve herein are not as significant as compare to the other cases discussed above.

In Fig. 6d, we show a log-log plot of the time variation of domain size ($\ell_C(t)$ *vs.* $t$) at various $k_a$ values for the evolution shown in Fig. S2. For comparison, we also plot the length scale for the simple ternary fluid mixture indicated by a blue line. As expected, we observe the smaller length scale for stiffer polymer chains (larger $k_a$) at a given time. Reason being, the stiffness in polymer chains results in the lack of the connectivity between A- and B-rich domains due to slower diffusion of chains, which subsequently affects the hydrodynamic transport. The crossover from the viscous ($\phi = 1$) to inertial hydrodynamic regime ($\phi = 2/3$), represented by the dashed lines is evident in our plot.

To understand the different morphologies of phase separating ternary mixtures, we further compare the radial distribution function ($g_{AB}(r)$) of a simple ternary fluid with the ternary fluids having one polymeric component, for three different polymer chain lengths ($L_p$) (shown by different symbol types), where, we consider the critical mixtures (A:B:C=1:1:1) at a fixed polymer chain stiffness ($k_a = 20$). Here we select $t = 200$ mainly for two reasons: first, all the system under consideration move into the scaling regimes by this time, and second, to avoid the finite-size effect (where domain sizes become comparable to the system size) that starts setting in for the simple ternary fluids beyond $t = 200$. The peak position of $g_{AB}(r)$ shifts towards lower $r$ with increasing chain length which confirms that the average domain sizes of A- or B-type beads become smaller with longer polymer chain.

Further, we plot the scaled correlation, $D(r,t)$ vs. $r/\ell_D$ at $t = 200$ to signify the domains of C-components in Fig. 7b. A clear deviation of the scaling functions from a master scaling function in different cases (represented by various symbol types) confirms the dependence of the scaling functions on the chain length. A similar comparison is also done in Fig. S3 (see SI) for three different polymer chain stiffness ($k_a = 5, 20$ and $60$) at a fixed polymer chain length ($L_p = 32$). In general, we distinctly observe the change in the average domain sizes of A- or B-type beads in the phase separating ternary (ABC) fluid by replacing one simple fluid component (C) with a polymeric fluid. The change in polymer chain composition ratio ($A:B:C$), chain length ($L_p$), chain stiffness ($k_a$) have fairly good effect on the domain morphology, and therefore, on the length scale.

## 4. Conclusions

In summary, we performed dissipative particle dynamics (DPD) simulations in $d = 3$, to study the evolution morphology of phase-separating ternary (ABC) fluid mixtures, where component C was a polymeric fluid. To place our results in proper context, we compared them with simple binary (AB) and ternary (ABC) fluid mixtures. We discussed our results mainly for two cases: (i) segregation of simple ternary (ABC) fluid mixtures with symmetric (1:1:1) and asymmetric (2:2:1 and 4:4:1) compositions and their comparison with simple binary (AB) fluid mixtures, and (ii) segregation of ternary (ABC) fluid mixtures with one polymeric component, where we explored the effect of polymeric fluid on the coarsening morphology by varying parameters such as, polymer concentration, polymer chain length, and polymer chain stiffness. In all the cases, results were obtained by averaging over five independent runs.

In case (i), we found that the scaling functions $C(r,t)$ and $S_C(k,t)$ appear to be independent of the composition ratios, while $D(r,t)$ and corresponding $S_D(k,t)$ exhibited a clear dependency on it. By calculating the RDF's of A-beads around B-beads, we observed that off-criticality leads to more segregated domains. More importantly, we have also accessed the crossover from viscous to inertial hydrodynamic regime ($\ell(t) \sim t^1 \rightarrow t^{2/3}$) in $d=3$. This is the first such observation for the phase separation dynamics in ternary (ABC) fluid mixtures so far. Similarly, in case (ii), we observed that the variation in composition ratio (A:B:C), polymer chain-length ($L_p$), and chain stiffness ($k_a$), did not change the scaling form of $C(r,t)$ and $S_C(k,t)$. However, $D(r,t)$ and its corresponding $S_D(k,t)$ are found to be clearly dependent on these parameters.

On further comparison of our results for various morphologies in both the cases, we noticed that the coarsening was affected more significantly by changing the composition of the component beads. In case of a ternary mixture with polymeric component, we again observed the crossover from the viscous hydrodynamic regime ($\phi = 1$) to the inertial hydrodynamic regime ($\phi = 2/3$). To the best of our knowledge, this is the first such significant observation for $d=3$ ternary fluid mixture with one polymeric component.

Polymeric fluids play a crucial role in various living things, *e.g.*, they are used in providing primary structural materials and participate in vital life processes. This model could be useful to understand the different morphologies in the system caused by the change in physical properties of the polymeric fluid. Our model could be useful to understand the industrial fluid/oil purification process where liquid-liquid phase separation (LLPS) (also termed as oiling-out) technique is widely utilized to produce fluid/oil with excellent physical properties.

In future work, we plan to investigate the polymerization controlled segregation of ternary fluid mixtures, where one or two components of the mixture are polymerizing. Further, we investigate such systems by varying the affinity of one component with other components of the mixture. We believe that our results will provide a generic framework for the analysis of simulations and experiments on the domain growth of ternary fluid mixtures and provoke a fresh interest in these little-explored systems.

## Acknowledgments

A.S. is grateful to CSIR, New Delhi for the financial support, A.S. is thankful to SCIS JNU, New Delhi, for providing the computational facilities. A.C. acknowledges the financial support from Grant No. BT/BI/03/004/2003(C) of Government of India, Ministry of Science and Technology, Department of Biotechnology, Bioinformatics division, and DST-PURSE grant to JNU by the Department of Science and Technology, New Delhi, India.## References

**Figures**

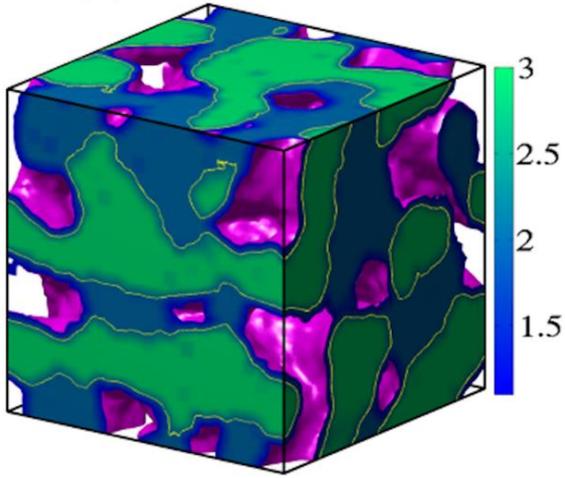
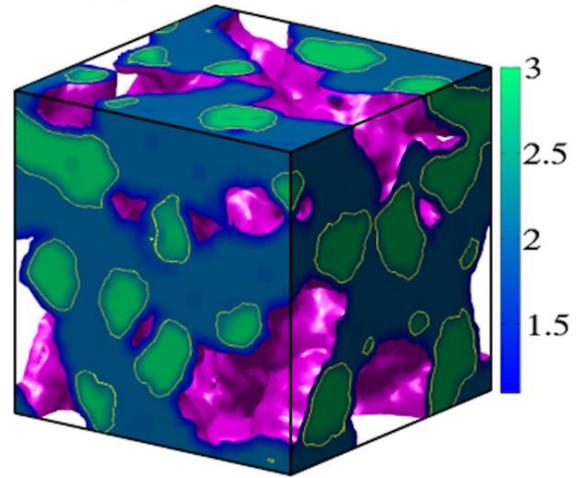
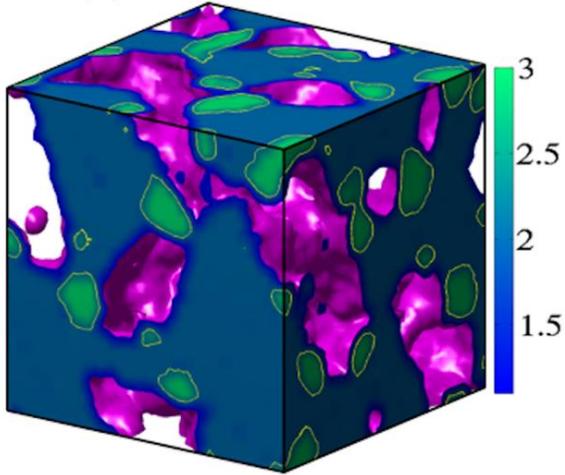
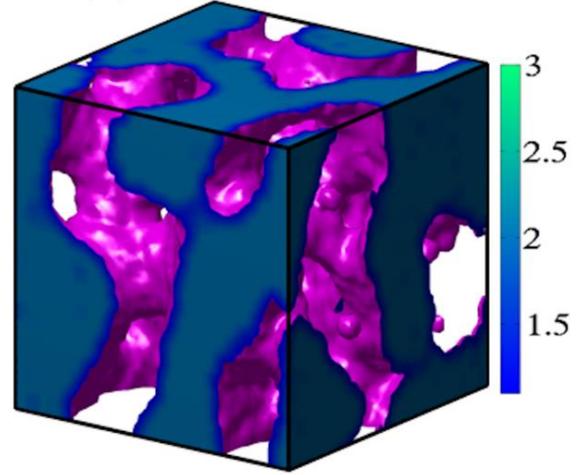

**Figure 1**

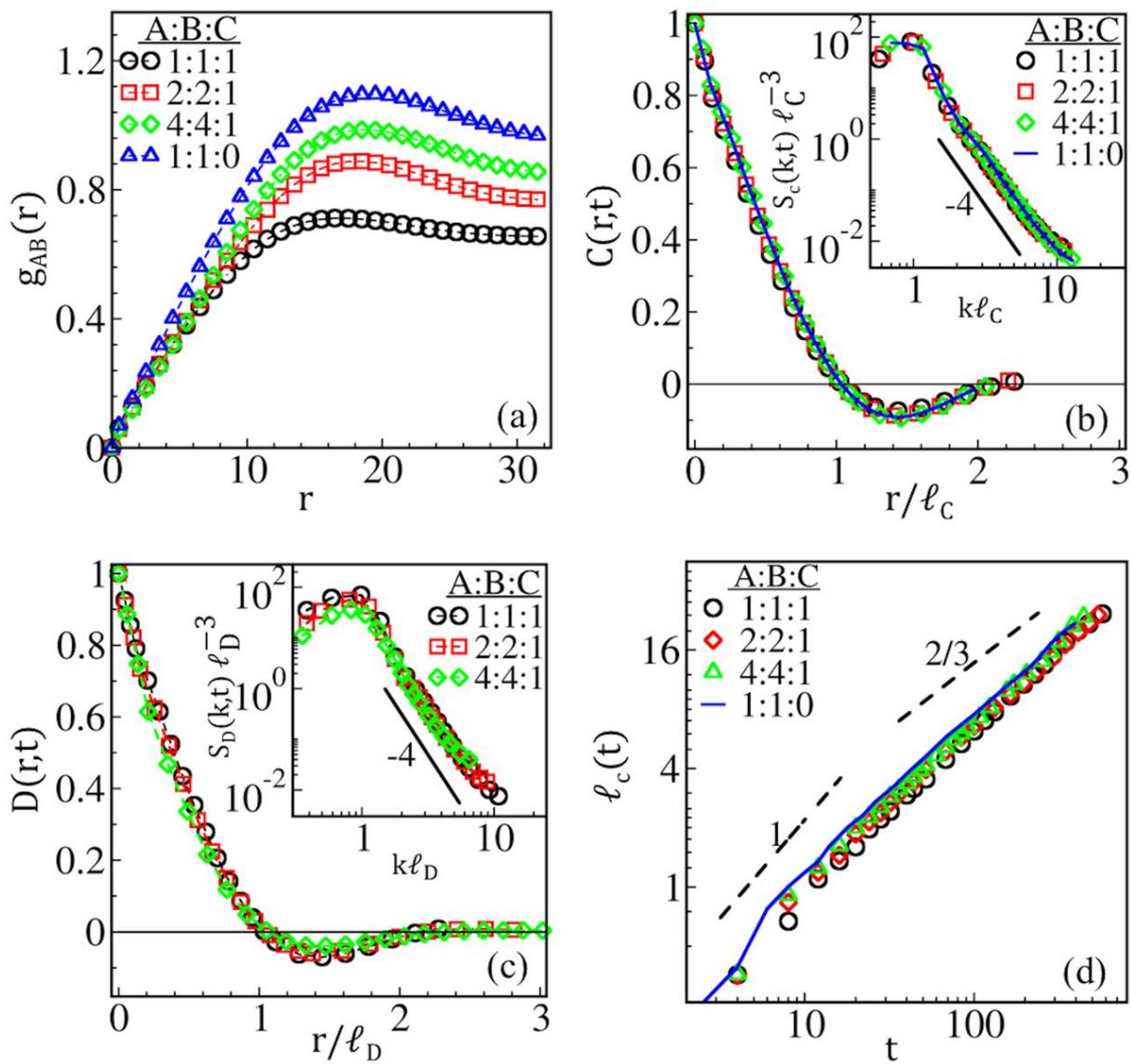

**Figure 2**

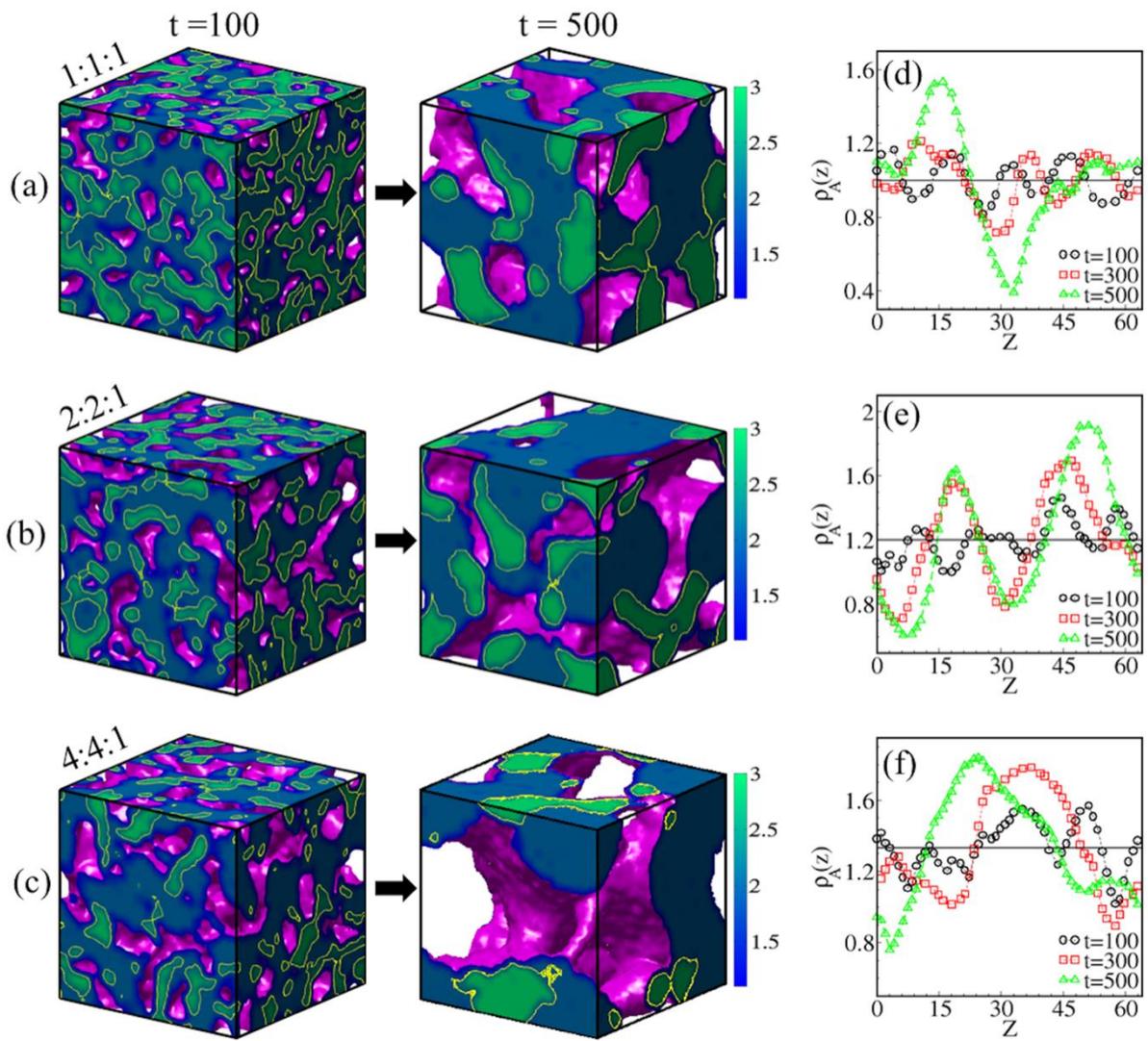

**Figure 3**

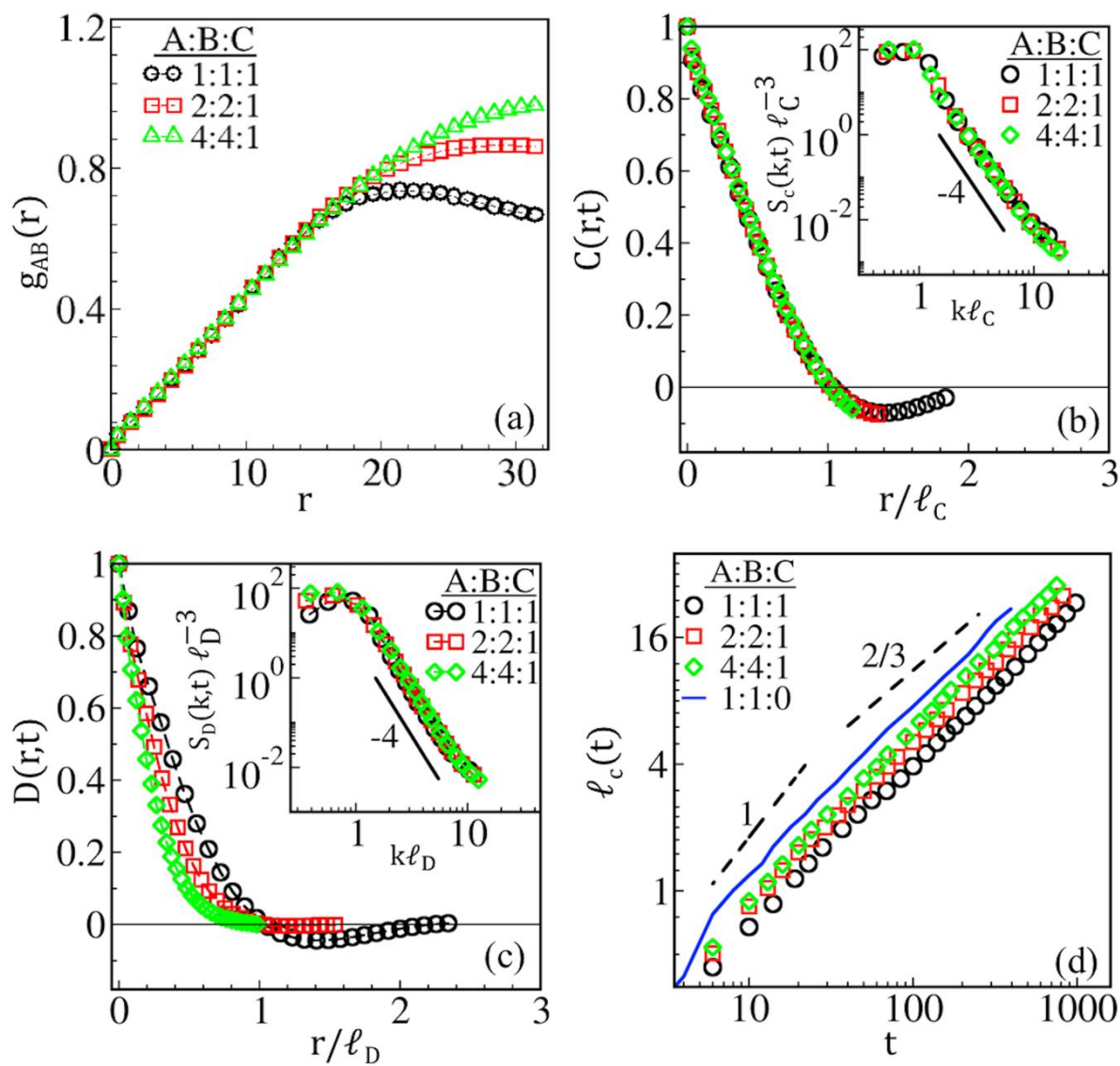

**Figure 4**

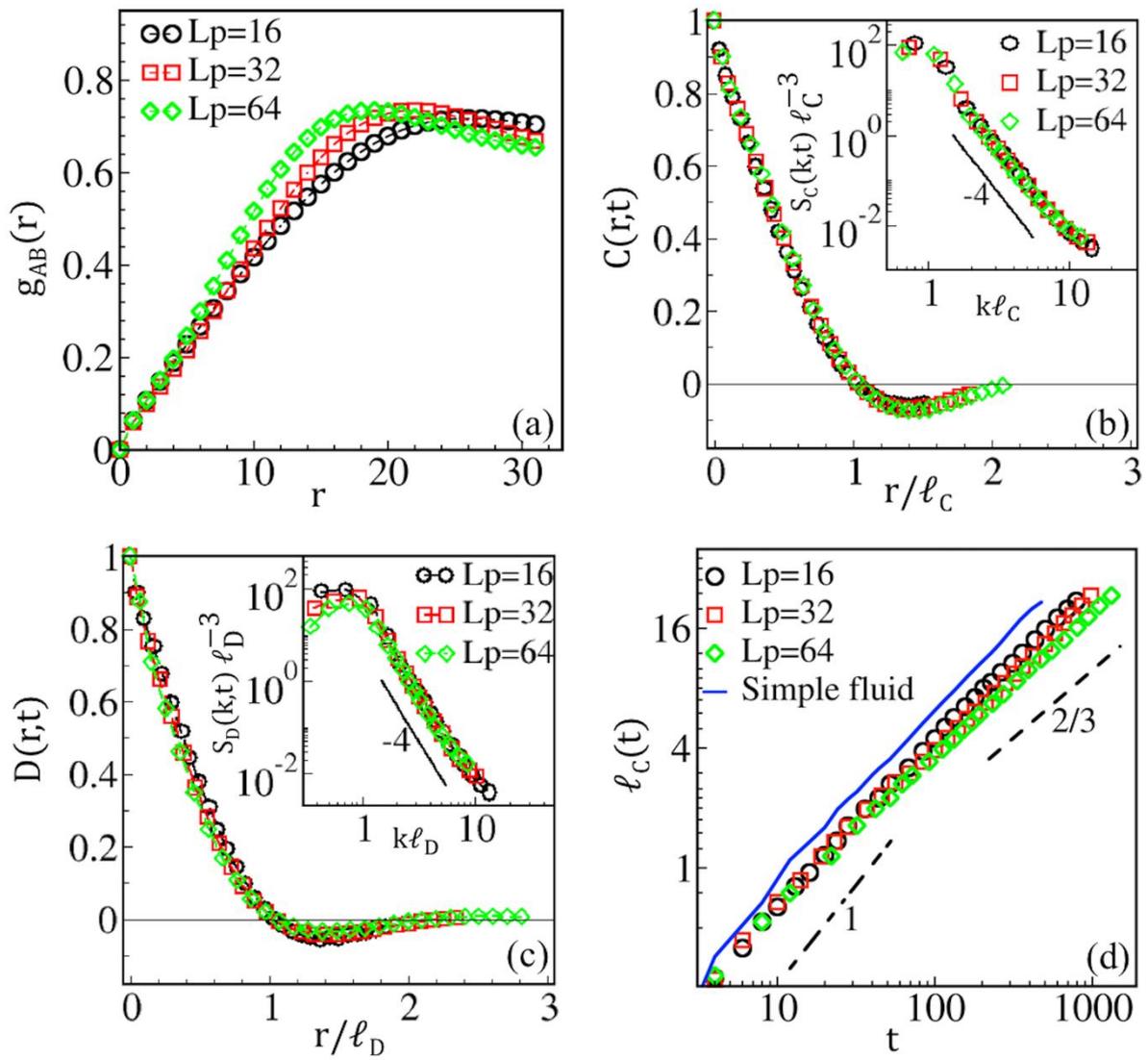

**Figure 5**

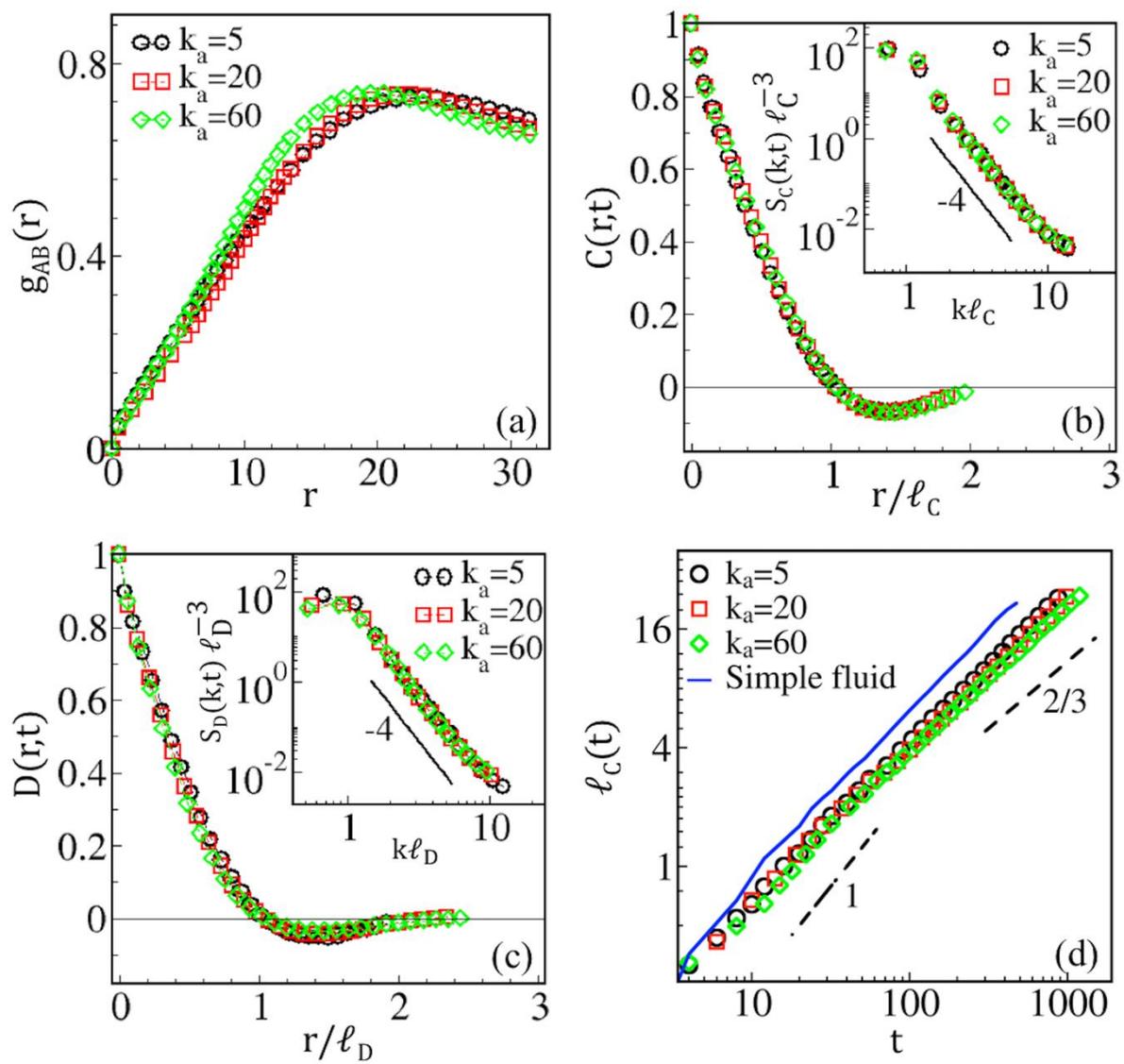

**Figure 6**

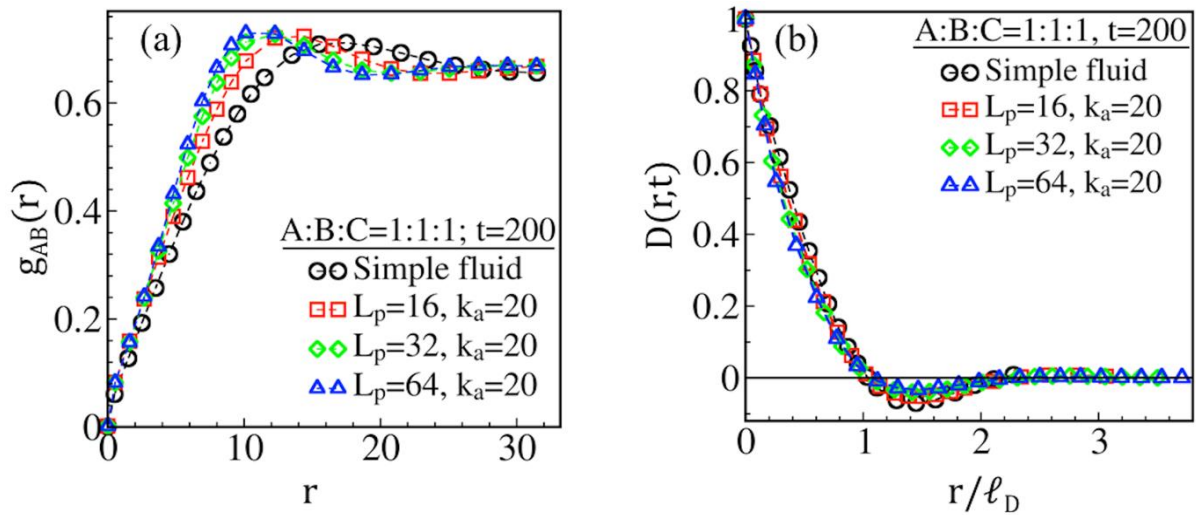

**Figure 7**

# Figure Captions

**Figure 1:** Snapshots (isosurfaces) of ternary mixtures in $d = 3$, at $t = 200$, for different ratios A:B:C (a) 1:1:1, (b) 2:2:1, (c) 4:4:1, and (d) 1:1:0. The open space, blue, and green patches represent A-rich, B-rich, and C-rich domains, respectively. The surface between A and B is shown in purple. Note that, only for the isosurface plot, we mapped $\psi(r,t)$ value in the range 1 to 3 (A ∈ (0, 1), B ∈ (1, 2), and C ∈ (2, 3)) to differentiate the colors of the interfaces (see the color bar).

**Figure 2:** (a) Comparison of the radial distribution functions ($g_{AB}(r)$) at $t = 200$, for different ratios A:B:C (represented by the symbols) for the evolution snapshots shown in Fig. 1. (b) The plot of $C(r,t)$ vs. $r/\ell_C$ at time $t = 200$. The solid line shows the scaled correlation function for a binary (AB) mixture. The inset plot is the corresponding structure factor, $S(k,t)\,\ell_C^{-3}$ vs. $k\ell_C$ for the same data sets on a logarithmic scale. Solid line of slope -4 represents the Porod's law, i.e., $S(k,t) \sim k^{-4}$ for $k \to \infty$. (c) Superposition of the data for $D(r,t)$ vs. $r/\ell_D$ and the inset picture is corresponding structure factor $S_D(k,t)\,\ell_D^{-3}$ vs. $k\ell_D$. (d) Log-log plot of the time dependence of the characteristic length scale $\ell_C(t)$. The solid line and open symbols represent the data sets for binary AB and ternary ABC fluids, respectively. The dashed lines of slopes 1 and 2/3 correspond to the expected growth regimes for $d = 3$ fluids. Details of our simulations are given in the text.

**Figure 3:** Domain morphology for ternary (ABC) fluids with a polymeric component (C) corresponding to three different ratios of A:B:C. The color mapping is same as in Fig. 1. The polymer chain stiffness ($k_a = 20$) and length ($L_p = 32$) are kept fixed for each plot. (a-c) show the evolution snapshots (isosurface) at $t = 100$ and $t = 500$ for $A:B:C = 1:1:1, 2:2:1$, and $4:4:1$, respectively. (d-f) represent the number density profile of A-type beads along the z-direction, at three different times $t = 100, 300$, and $500$ (shown by the symbols) for the respective ratios of A, B, and C, respectively.

**Figure 4:** (a) Comparison of $g_{AB}$ at time $t = 500$ for three different ratios of A, B, and C, indicated by the symbol types. (b) Comparison of $C(r,t)$ vs. $r/\ell_C$ at time $t = 500$. The inset diagram is the corresponding structure factor plot, i.e., $S_C(k,t)\ell_C^{-3}$ vs. $k\ell_C$ with Porod tail behavior shown by the solid line of slope $-4$. (c) Superposition of the data for $D(r,t)$ vs. $r/\ell_D$ (inset plot is $S_D(k,t)\ell_D^{-3}$ vs. $\ell_D$ ) at $t = 500$. (d) Time dependence of domain length scale $\ell_C(t)$ vs. $t$ on a log-log scale corresponding to the evolutions shown in Fig. 3. Again, the dashed lines of slope 1 and 2/3 represent the expected growth regimes for segregating ternary fluids in three-dimensions. The solid blue line represents the length scale of simple ternary fluid and is plotted for the comparison with other length scales.

**Figure 5:** Results when polymer chain stiffness ($k_a = 20$) and composition ratio ($A:B:C = 1:1:1$) are kept constant. (a) Comparison of $g_{AB}$ at $t = 500$ for three different polymer lengths (shown by different symbols). (b) Comparison of $C(r,t)$ vs. $r/\ell_C$ at time $t = 500$. Symbols show three different polymer lengths as mentioned. The inset diagram presents $S_C(k,t)\ell_C^{-3}$ vs. $k\ell_C$ with Porod tail behavior shown by the solid line of slope $-4$. (c) Superposition of the data for $D(r,t)$ vs. $r/\ell_D$ (inset plot is $S_D(k,t)\ell_D^{-3}$ vs. $k\ell_D$). (d) Time dependence of domain length scale $\ell_C(t)$ vs. $t$ on a log-log scale corresponding to the evolutions in Fig. S1. The solid blue

line is plotted for the comparison with the length scale of simple ternary fluid. The dashed lines of slope 1 and 2/3 represent the expected growth regimes for phase separating three-dimensional ternary fluids.

**Figure 6:** (a) Comparison of $g_{AB}$ at $t = 500$ for three different $k_a$ values shown by different symbol types. (b) Comparison of $C(r,t)$ vs. $r/\ell_C$ at $t = 500$. The inset diagram is $S_C(k,t)\ell_C^{-3}$ vs. $k\ell_C$ with Porod tail (slope $-4$). (c) Superposition of the data for $D(r,t)$ vs. $r/\ell_D$ (see inset for $S_D(k,t)\ell_D^{-3}$ vs. $k\ell_D$) at $t = 500$. (d) Time dependence of average domain size $\ell_C(t)$ vs. $t$ on a log-log scale. Here, we have kept the polymer chain length ($L_p = 32$) and the composition ratio (A: B: C $= 1:1:1$) fixed corresponding to the evolutions in Fig. S2. Dashed lines of slope 1 and 2/3 represent the expected growth regimes for segregating three-dimensional ternary fluids and blue solid line shows the length scale of simple ternary fluids.

**Figure 7:** (a) Comparison of RDF ($g_{AB}$) of simple ternary fluids and ternary fluids with one polymeric component for three different chain lengths at $t = 200$ by keeping chain stiffness fixed at $k_a = 20$. Here, we consider only the critical ternary fluids ($A: B: C = 1:1:1$) for comparison. (b) Comparison of $D(r,t)$ vs. $r/\ell_D$ at $t = 200$ also shows a clear deviation of the scaling functions from each other in different cases as depicted by the various symbol types.

**Role of a polymeric component in the phase separation of ternary fluid mixtures: A dissipative particle dynamics study**

Amrita Singh[a], Anirban Chakraborti[a], and Awaneesh Singh[b]

[a] School of Computational and Integrative Sciences, Jawaharlal Nehru University, New Delhi-110067, India.

[b] Department of Physics, Institute of Chemical Technology, Mumbai-400019, India. E-mail: awaneesh11@gmail.com

**Supplementary Information**

In Fig. S1, we present the isosurface of the evolving system for chain length $L_p = 16$, and 64 at two different times $t = 100$ and 500, whereas, the isosurface for $L_p = 32$ is already shown in Fig. 3a. After the quench, as expected for symmetric mixtures, we observe the formation of A-rich (open space), B-rich (blue patch), and C-rich (green patch) blobs with all three types of interfaces, namely AB, BC, and CA present. We plot the number density profiles of A-beads ($\rho_A(z)$) along the z-direction corresponding to $L_p = 16$ (in Fig. S1c), $L_p = 32$ (in Fig. 3d), and $L_p = 64$ (in Fig. S1d) at three different times ($t = 100, 300$, and 500). The spreading of curves with time evidently supports the coarsening phenomenon. The density profiles (e.g., at $t = 500$) clearly demonstrate that the average domain size is larger for shorter polymer chain length ($L_p = 16$) as compared to that for other two chain lengths $L_p = 32$, and 64, respectively.

Next, we present the evolution snapshots (isosurfaces) at $t = 100$ and 500 for $k_a = 5$ (Fig. S2a) and 60 (Fig. S2b), respectively. The domain sizes are clearly larger for the system with $k_a = 5$ as compared to that for $k_a = 60$, at a given instant of time. This is expected as the polymer chains with a smaller value of $k_a$ will be more flexible to move. In such systems, A and B- beads can be accommodated in their domains in a faster and easier way as compared to the system with larger $k_a$ (more rigid/less flexible polymers). This is also apparent from Figs. S2c and S2d where, we plot the number density profile of A-type beads along the z-direction for $k_a = 5$ and 60 at three different times as mentioned and indicated by the symbols.

In Fig. S3a, we compare $g_{AB}(r)$ of a critical simple ternary fluid (A:B:C=1:1:1) with the critical polymer-containing ternary fluids, for three different stiffness ($k_a$) values at $t = 200$. The shifting of peak position of $g_{AB}(r)$ towards lower $r$ confirms that the average domain sizes of A- or B-type beads become smaller with increasing stiffness of the polymer chains. Further, we plot the scaled correlation, $D(r,t)$ vs. $r/\ell_D$ at $t = 200$ to signify the domains of C-components in Fig. S3b. A clear deviation of the scaling functions from each other in different cases (represented by various symbol types) confirms the dependence of scaling functions on the chain stiffness. Overall, the average domain size of A- or B-type beads changes clearly by replacing one simple fluid component (C) with a polymeric component in the phase separating ternary (ABC) fluid mixtures. The change in polymer chain stiffness also have fairly good effect on the domain size.

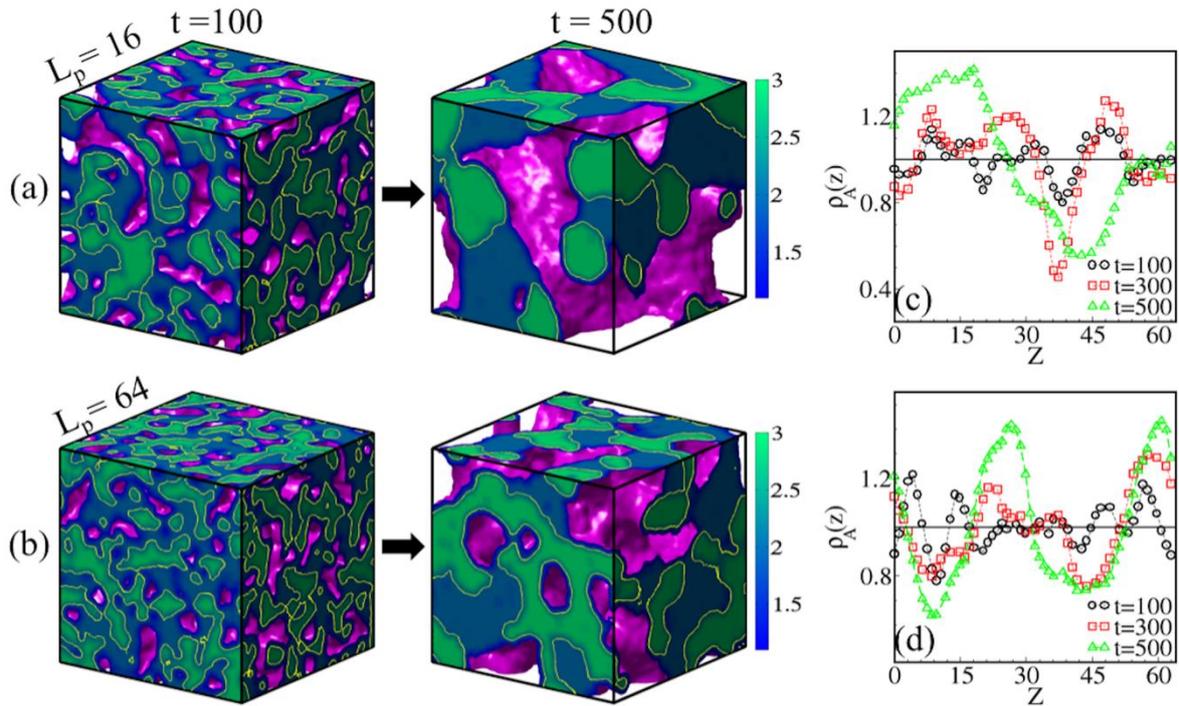

**Figure S1:** Evolution morphology of the ternary fluids corresponding to two different lengths of polymers (C-type molecules) for the case where $A:B:C = 1:1:1$ and $k_a = 20$. (a-b) represent evolution snapshots (isosurfaces) at $t = 100$ and $500$ for $L_p = 16$ and $64$ respectively. (c-d) Number density profile of A-type molecules along the z-direction, at three different times $t = 100$, $300$, and $500$ represented by the symbols for the respective chain lengths. $L_p = 32$ case is already shown in Fig. 3a and 3d.

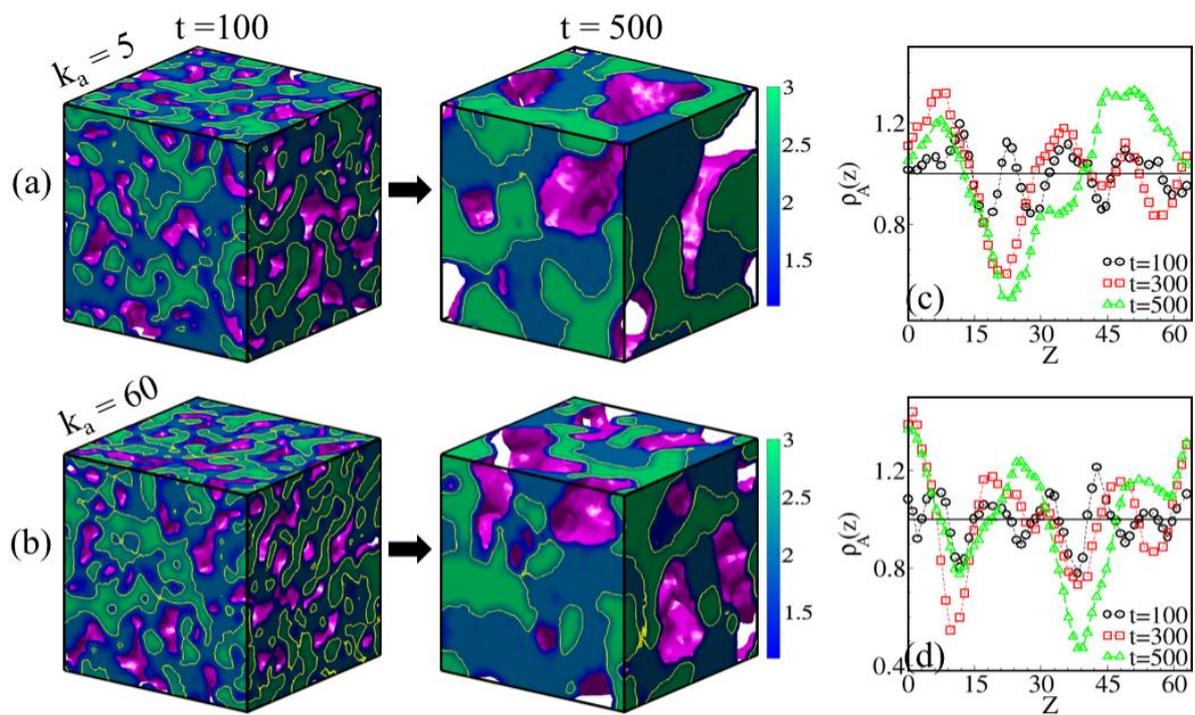

**Figure S2:** Evolution morphology of the ternary fluids (with one polymeric component) for $k_a = 5$ and 60 for the case where $A:B:C = 1:1:1$ and $L_p = 32$, in each case. (a-b) Display evolution snapshots at $t = 100$ and 500 for $k_a = 5$ and 60, respectively. (c-d) Number density profiles of A-type molecules at three different times, $t = 100, 300,$ and 500 shown by the symbols for the respective $k_a$ values of the polymers. For the case $k_a = 20$, Fig. 3a and 3d is referred.

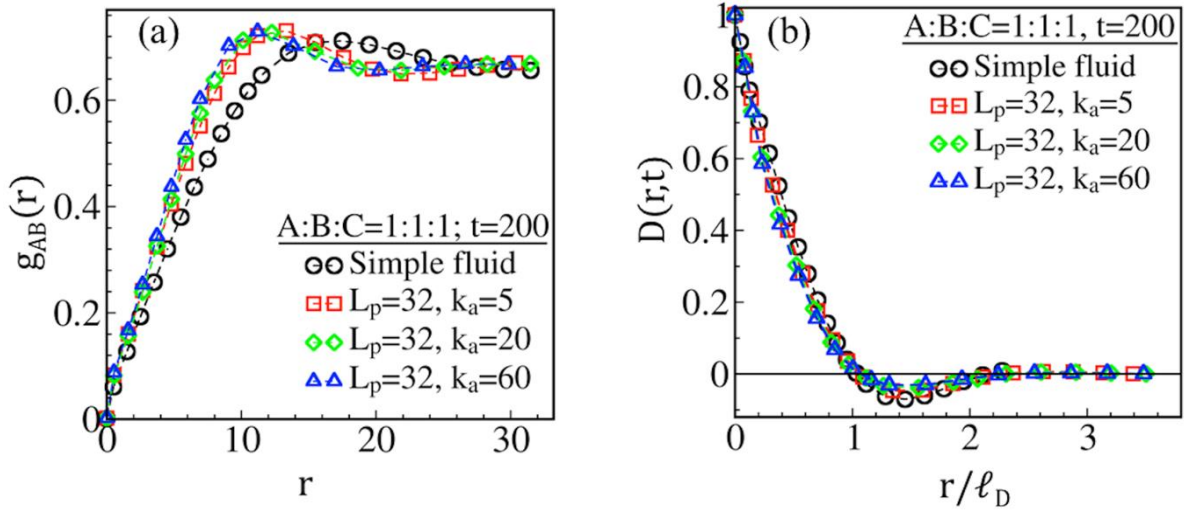

**Figure S3:** (a) Comparison of $g_{AB}$ of simple fluid mixture and mixtures with one polymeric component for three different $k_a$ values (shown by different symbol types) at $t = 200$. Here, we consider the critical mixtures (A:B:C=1:1:1) only. (b) Comparison of $D(r,t)$ vs. $r/\ell_D$ at $t = 200$ shows a clear deviation of the scaling functions from each other in different cases represented by the various symbol types.